\documentclass[rmp,showpacs,twocolumn,superscriptaddress,floatfix]{revtex4}

\usepackage{dcolumn,graphicx,amsmath,amssymb,txfonts}

\begin{document}

\title{Radial Structure of the Internet}

\author{Petter Holme}
\affiliation{Department of Computer Science, University of New Mexico,
  Albuquerque, NM 87131, U.S.A.}

\author{Josh Karlin}
\affiliation{Department of Computer Science, University of New Mexico,
  Albuquerque, NM 87131, U.S.A.}

\author{Stephanie Forrest}
\affiliation{Department of Computer Science, University of New Mexico,
  Albuquerque, NM 87131, U.S.A.}
\affiliation{Santa Fe Institute, 1399 Hyde Park Road, Santa Fe, NM
  87501, U.S.A.}

\begin{abstract}
The structure of the Internet at the Autonomous System (AS) level has
been studied by both the Physics and Computer Science communities. We
extend this work to include features of the core and the periphery,
taking a radial perspective on AS network structure.  New methods for
plotting AS data are described, and they are used to analyze data sets
that have been extended to contain edges missing from earlier
collections. In particular, the average distance from one vertex to
the rest of the network is used as the baseline metric for
investigating radial structure.  Common vertex-specific quantities are
plotted against this metric to reveal distinctive characteristics of
central and peripheral vertices. Two data sets are analyzed using
these measures as well as two common generative models
(Barab\'{a}si-Albert and Inet).  We find a clear distinction between
the highly connected core and a sparse periphery.  We also find that
the periphery has a more complex structure than that predicted by
degree distribution or the two generative models.
\end{abstract}

\pacs{89.20.Hh,89.75.Fb, 89.75.Hc}

\maketitle

\section{Introduction}

Since the turn of the century there has been increasing interest in the
statistical study of networks~\cite{ba:rev,doromen:book,mejn:rev},
stimulated in large part by the availability of large-scale network
data sets.  One network of great interest is the
Internet~\cite{vesp:inet}. The Internet is intriguing because its
complexity and size preclude comprehensive study. It is comprised of
millions of individual end-nodes connected to tens of thousands of ISPs
whose relationships are continually in flux and only partially
observable.  One way to cope with these complexities is by analyzing a
single scale of Internet data, for example, a local office network of
computers and their inter-connections; a network of email address book
contacts; the network formed by URL links on the World Wide Web; or the
interdomain (Autonomous System) level of the Internet. This paper is
concerned with the last of these examples---the AS graph. The vertices
in the graph are themselves computer networks; roughly speaking an AS is
an independently operated network or set of networks owned by a single entity. Edges represent pairs of ASs that can directly
communicate.

A major finding of earlier AS studies is that node degree (number of
links to other ASs) has a power law distribution~\cite{f3}. The degree
distribution is, however, not the only structure that affects 
Internet dynamics~\cite{hot:inet}. In this paper we investigate
higher-order (beyond the degree distribution) network structures that
also impact network dynamics. We analyze the AS graph using methods
that are appropriate for networks with a clear hierarchical
organization~\cite{vesp:inet,ala:hier}. In particular, we study
network quantities as a function of the average distance to other
vertices. This approach allows us to separate vertices of different
hierarchical levels, in a radial fashion, ranging from central (in the
sense of the closeness centrality~\cite{sab:clo}) to peripheral
vertices. This is, furthermore, a way to dissolve how clearly
separated the core and the periphery are. Most analysis methods
developed by physicists (degree frequencies, correlations, etc.)\ are
based on quantities averaged over the whole network and do not take a
hierarchical partitioning into account~\cite{vesp:inet}. Studies by
computer scientists, on the other hand, assume a division of the AS
level Internet into hierarchical levels~\cite{rex:infer}.  We will
argue that the observed AS level networks do have pronounced
core-periphery dichotomy but that the periphery has more structure
than previously thought.

\section{Networks}

This section briefly reviews the organization of the
AS-level Internet and describes how we obtained our data sets. We also
describe the network models to which we compare our observed
data. These models include one randomization scheme that samples random
networks with the same set of degrees as the original networks, and the generative BA and Inet models. Technically all three models are
null-models, but to contrast the randomized networks (having $N$
degrees of freedom) with the generative models (having only a few
degrees of freedom) we reserve the term null-model to the former.

The data are represented as a network $G=(V,E)$ where $V$ is a set
of $N$ vertices (ASs) and $E$ is a set of $M$ undirected edges 
(connections between ASs).
The Internet is currently composed of roughly $22,000$ individual
networks known as Autonomous Systems.  Each of these systems peer with
a (usually small) set of ASs to form a connected network.  The
protocol used to establish peering sessions and discover routes to
distant ASs is called the Border Gateway Protocol (BGP).  Two typical
peering relationships are: customer-provider in which the provider
provides connectivity to the rest of the Internet for the customer;
and peer-peer in which the peering ASs transfer traffic between their
respective customers. The extreme core of the network, the Tier-1 ASs, have many peer-peer and
customer links but no providers.  Nodes closer to the periphery of the
network have fewer customers and peers but more providers.

\subsection{AS networks}

We analyze four real-world data sets (that is, data sets collected
using observed network data rather than simulated networks that are
generated synthetically), of which two are original. The first two are
well-known and well-studied~\cite{mich:as} dating from 2002 and the
second two data sets are recent, inferred from 2006 data. The first
graph in each pair consists of edges learned solely from dumps of
router state, known as Routing Information Bases (RIBs)
(\url{http://www.routeviews.org/data.html}).  RIBs are a standard
source of AS connectivity data.  The second graph in each pair
contains RIB information augmented with edges derived from other
sources (such as routing registries, looking glass servers, and update
messages) which produces a more accurate representation of the real
network.  The additional sources are described below.

\subsubsection{Obtaining RIBs from Route Views}

BGP routers store the most recent AS path for each IP block (prefix)
announced by its peers.  These data are stored in the router's RIB,
and periodic RIB dumps from a large number of voluntary sources are
available from Route Views (\url{http://www.routeviews.org}). Each RIB
represents a static snapshot of all routes available to the router
from which it was obtained.  Since BGP only disseminates each router's
best path, and this value is dynamic as links go up and down, a
sizable portion of the network is hidden from each router.  In order
to obtain a more complete topology, common practice is to take the
union of the relationships found in a large number of RIB samples.
From the samples, AS relationships are then inferred from the routing
paths.  A path is comprised of connected ASs and therefore each pair
of adjacent ASs in a path corresponds to an edge in the graph.

The 2002 graph taken from a single RIB (RIB '02) was inferred from Route
Views on May 15th of 2002.  We constructed the 2006 RIB graph (RIB '06)  
from the Route Views RIB on May 16th of 2006. The RIB '02 graph has
$N=13233$ and $M=27724$ while RIB '06 has $N=22403$ and $M=46343$.

\subsubsection{Extending the RIB Dataset}
\label{sec:extended}

There are other sources of AS connectivity data besides Route Views.
RIPE (\url{http://www.ripe.net}) has data collected from additional RIBs
beyond those contained in the Route Views data.  Peering information
is directly available for a small number of ASs that are participating
Looking Glass (\url{http://www.traceroute.org}) routers.  Finally, some
ASs register their peering relationships in regional registries such
as RIPE. The extended 2002 AS graph (AS '02) was constructed using
inferred topologies from all three of these sources, together with the
original Route Views data.

RIB data represent a brief snapshot of routing state.  There are many
paths that a router sees only briefly, and the chances of capturing
all of them from just a few RIB dumps is unlikely.  In the extended
AS-graph of 2006 (AS '06), we augmented the Route Views RIB data with
all of the paths found in BGP update messages for the entire month of
April 2006 from both Route Views and RIPE.  This gives a more
complete picture over time, although it is still biased by the limited
number of routers from which the data were collected.

The extended 2002 AS-graph (AS '02) has $N=13579$, $M=37448$ and the
corresponding 2006 network (AS '06) has $N=22688$ $M=62637$. Thus the
extended data sets have $35\%$ (2002) and $67\%$ (2006) more edges than
their RIB counterparts.

\subsection{Null-model networks}

We are interested in network structure beyond degree distribution, so
we compare our AS network data against a null model
with the same degree distribution.  Our null model is a random network
constrained to have the same set of degrees as the original network.
By comparing results for the observed networks with the same quantities
for the null model, we can observe additional network structure if it
exists.  The standard way to sample such networks is by randomizing
the original network with stochastic rewiring of the edges (see
Ref.~\cite{gale:rew} for an early example). In our implementation we
create a new random network by enumerating the edges $E$ of the
original graph, and for each edge $(i,j)$ we are:
\begin{enumerate}
\item Choosing another edge $(i',j')$ randomly and replacing $(i,j)$
  and $(i',j')$ with $(i,j')$ and $(i',j)$. If this creates a
  multi- or self-edge, then we are reverting to the original edges  $(i,j)$ and
  $(i',j')$, and repeating with a new  $(i',j')$.
\item \label{step:rew3} Choosing two edges
  $(i_1,j_1)$ and $(i_2,j_2)$ and replacing them along with $(i,j')$ by
  $(i_1,j')$, $(i,j_2)$ and $(i_2,j_1)$.
\end{enumerate}
Step \ref{step:rew3} guarantees ergodicity of the
sampling~\cite{roberts:mcmc}, i.e.\ that one can go between any pair of graphs
with a given set of degrees by successive edge-rewirings.

\subsection{Generative network models}

In addition to the observed (inferred from data) and null-model networks described above,
we also study networks produced according to two previously proposed
network-generation schemes~\cite{ba:model,inet}. The first is the well-known the
Barab\'{a}si-Albert preferential attachment model~\cite{ba:model}.
The second, known as the Inet model (version 3.0)~\cite{inet}, is
more complex and designed specifically for creating networks with AS
graph properties.

\subsubsection{Barab\'{a}si-Albert model}

The Barab\'{a}si-Albert (BA) model is a general growth model for
producing networks with power-law degree distributions
Ref.~\cite{ba:model}. Vertices and edges are iteratively added to the
network according to a preferential attachment rule, which ensures
that a power-law degree distribution emerges.  

More precisely, the initial configuration consists of $m$ isolated
vertices. From this configuration the network is iteratively grown.
At each time step one vertex is added together with $m$ edges leading
out from the new vertex. The edges are attached to vertices in the
graph such that:
\begin{enumerate}
\item The probability of attaching to a vertex $i$ is proportional to
  $k(i)$.
\item No multiple edges, or self-edges, are formed.
\end{enumerate}
This procedure produces a network which has, 
in the $N\rightarrow\infty$ limit, a degree
distribution $P(k)\sim k^{-3}$ for $k\geq m$, and $P(k)= 0$ for $k<m$.

Because the BA model has only one integer parameter it is not very
flexible at fitting data. We use $m=3$ to make the average degree
as similar to the AS networks as possible. Other preferential
attachment models (e.g., Ref.~\cite{pas:inet}), can model the average
degree and slope of the degree distribution more closely. Such
improvements, we believe, are unlikely to change the conclusions drawn from the
original BA model.

\subsubsection{Inet model}

The Inet model~\cite{inet} is less general than BA's. Its objective is
to regenerate the AS graph as accurately as possible rather than to focus
on a single mechanism to create and explain scale-free networks.  The
scheme is rather detailed and we only sketch its strategy here.  Starting with $N$ vertices, Inet first generates random
numbers that represent the final degree of the vertices such that
the degree distribution matches the observed distribution of the
AS-graph as closely as possible. This means that the low-degree end of
the distribution is more accurately modeled by Inet than the BA model
because the BA model will not produce a vertex with degree less than $m$.
In the real AS-graph there are a considerable fraction of
degree-one vertices. After the degrees are assigned to the vertices,
edges are added in such a way that the degree-degree correlation
properties of the original AS-graph is matched as closely as
possible.

A more detailed explanation of this procedure and its rationale are
given in Ref.~\cite{inet}. We use Inet's default parameter settings,
except $N$ which we extracted from our datasets, producing an average
degree that is approximately six.

\section{Numerical results}

\begin{figure}
  \resizebox*{0.9 \linewidth}{!}{\includegraphics{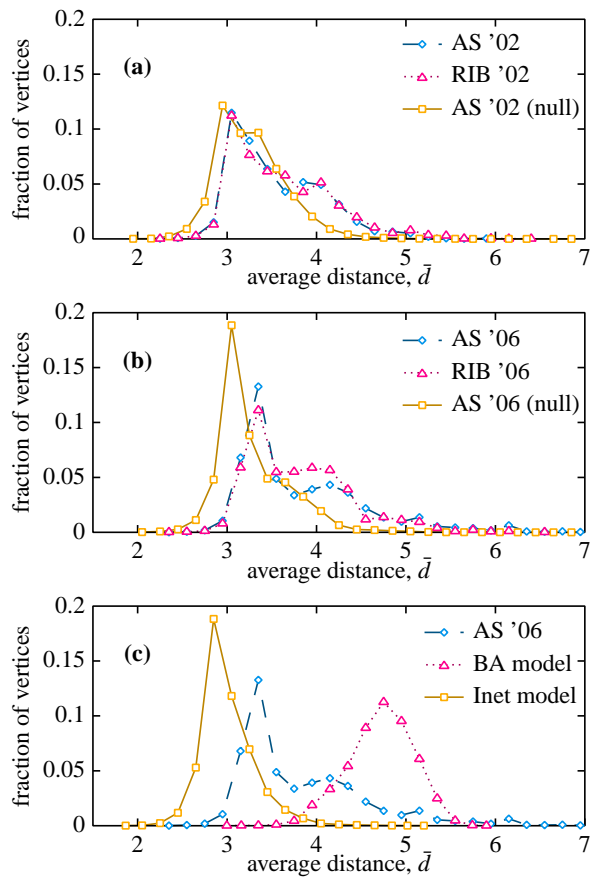}}
  \caption{ Normalized histograms of vertices with a specific average
    distance $\bar{d}$ to the rest of the vertices. (a) shows curves for the
    Oregon Route Views data (RIB '02), extended data (AS '02), and values for
    random networks with the same degree sequences as AS '02. (b)
    displays curves for the Oregon Route Views data (RIB '06), extended
    data (AS '06), as well as randomized
    networks with the degree sequence of AS '06. (c) shows the same AS
    '06 curve as (b) along with the BA and Inet model results for
    parameter values as close as possible to those of the AS '06
    network. 100 averages were used for the null-model curves in (a)
    and (b) as well as the model networks in (c). Lines are guides for
    the eyes. The error-bars represent standard error (the point
    symbols are often larger than the error bars).
  }
  \label{fig:density}
\end{figure}

In this section we present the numerical results of our analysis. We
first discuss the average distance metric we use for displaying
network properties with a radial perspective. Then we define and
present the results for each network structural measure as a function
of the average distance to other vertices.

Let $d(i,j)$ denote the graph distance between two vertices $i$ and
$j$---the number of edges in the shortest path between $i$ and $j$. A
simple measure for how peripheral a vertex is in the network is its
\textit{eccentricity}---the distance to the most distant vertex,
$\max_{j\in V} d(i,j)$~\cite{harary}. Eccentricity is thus an extremal
property of the network and is determined by a small fraction of
vertices. To reflect the typical path length of a vertex we rank
vertices according to an average property of the vertex.  The
average property corresponding to eccentricity is the average distance
from one vertex to all of the others:
\begin{equation}\label{eq:dist}
  \bar{d}(i)=\frac{1}{N-1}\sum_j d(i,j) ,
\end{equation}
where the sum is over all vertices, except $i$, in $V$. We note that
the reciprocal value of $\bar{d}(i)$, the \textit{closeness
  centrality}, is a common measure for centrality in social network
studies~\cite{sab:clo,harary}. Average distance is a more intuitive measure in this context---$\bar{d}(i)\approx 2$ means that $i$ is on average
two hops away from other vertices, whereas the closeness value $0.5$
does not have such a direct interpretation.

Another way to study eccentricity is by iteratively removing vertices
of low-degree to construct a sequence of $k$-cores (subgraphs in which
all vertices have degree $\geq k$)~\cite{rex:infer,vesp:kcore}. We used the average distance metric instead because it measures
separation of vertices---i.e.\ the values on the x-axis
are not only integers as for the eccentricity. Further, because it is
a global measure (in the sense that the entire network topology
affects $\bar{d}(i)$ for every $i$) it is likely more robust to errors
in the input data.

\subsection{Radial vertex density}
We first plot the fraction of
vertices as a function of $\bar{d}$. Fig.~\ref{fig:density} shows the
distribution of $\bar{d}$ for our data sets and the model AS
graphs. The observed networks produce graphs that are far from smooth,
unimodal distributions. Instead they have one peak close to
$\bar{d}=3$, a smaller peak around $\bar{d}=4$, and for the 2006 data,
a third peak near $\bar{d}=5$. The difference between the RIB-only and
the extended datasets is small, except around the second peak
in Fig.~\ref{fig:density}(b) which is higher in the RIB-only data. The
null-model curves are much more unimodal, although
they do not follow a simple, smooth functional form. Such a unimodal
form could be a result of the averaging of many null-model
curves, but the observation holds even if single realizations of the
randomization are plotted (data not shown). Thus, the observed AS graph is less homogeneous than what we would predict by considering only vertex degree.

We interpret the two peaks as an effect of the hierarchical
organization of the Internet. The core (Tier-1 providers and other
large ISPs) is in the low-$\bar{d}$ tail, the $\bar{d}=3$ peak are
vertices directly connected to the core, and the $\bar{d}=4$ peak are
vertices whose closest neighbors are in the $\bar{d}=3$ peak. This
explains the approximately integer distance between the peaks.
Determining the edge relationship between the peaks (customer-provider or peer-peer) is a difficult problem~\cite{rex:infer} however we believe that they are likely to be from customers to providers as ASs generally only have peer-peer edges with networks of equal class.  
The Tier-1 ASs that do not have any providers and are thus most core (AS numbers 209, 701,
1239, 1668, 2914, 3356, 3549, 3561, 6461 and 7018 in our data sets)
have an average $\bar{d}=2.35\pm 0.03$ in the AS '02 data and
$\bar{d}=2.41\pm 0.03$ in the AS '06 data, and are thus in the center
of the network (left of the most central peak).  Thus, the Tier-1 ASs are in 
the extreme low end of the $\bar{d}$-spectrum.

Results for the BA and Inet model networks are shown in
Fig.~\ref{fig:density}(c). The Inet model has a peak to the left of
the middle of the range of distances, but no second or third peak. 
The BA model matches the observed network even less accurately---its peak is at a relatively high $\bar{d}$ value.

\begin{figure}
  \resizebox*{0.9 \linewidth}{!}{\includegraphics{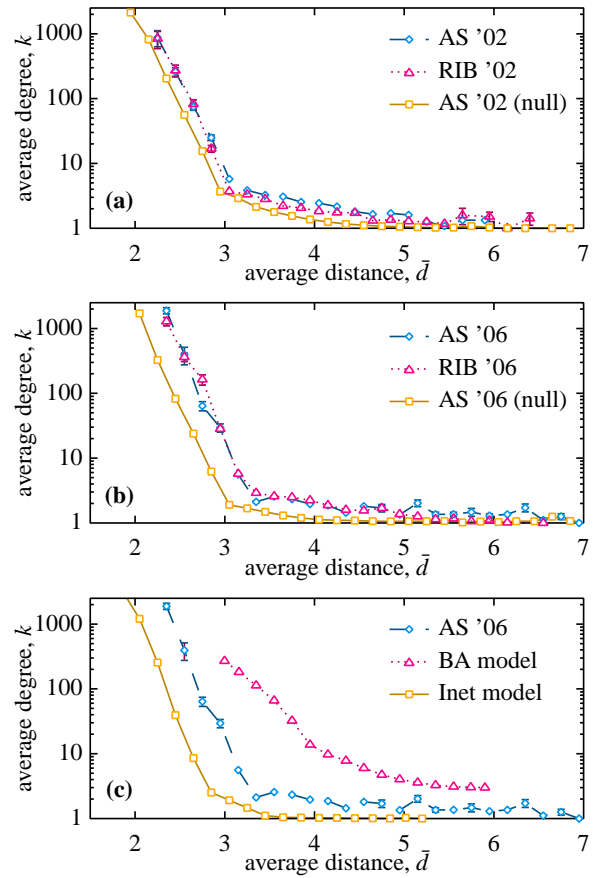}}
  \caption{ Degree $k$ as a function of the average distance
    $\bar{d}$. The panels and symbols represent the same
    data sets as in Fig.~\ref{fig:density}.}
  \label{fig:degree}
\end{figure}

\subsection{Degree}

Degree distribution is now a classical quantity in the study of the
Internet topology. Ref.~\cite{f3} reports a highly
skewed distribution of degree, fitting well to a power-law with an
exponent around $2.2$. Since this finding, the degree distribution has
become a core component in models of the AS graph---both the BA and
Inet models as well as others~\cite{fkp:model,ahs:model,meina:pow}
create networks with power-law degree distributions. One
interpretation of degree is that it is a local centrality
measure~\cite{harary}. Further,
different measures of centrality are known to be highly
correlated~\cite{centr:keiko,lee:corr,our:attack} so one can expect
the average degree $k$ to be a decreasing function of the average
distance $\bar{d}$.

Figure~\ref{fig:degree} confirms this prediction for both the observed
and model networks. In Fig.~\ref{fig:degree}(a) and (b) we observe
that the $k(\bar{d})$-curves decrease dramatically until the
approximate location of the first peak in the distribution plots
Fig.~\ref{fig:density}(a) and (b). Therefore, $\bar{d}$ identifies a
natural border between the core vertices of high-degree and low
average distance, and the sparsely connected periphery. The observed graphs,
however, have higher degree in the periphery compared to the
null-model curves. This suggests that the network periphery
may have more complex wiring topology than that is predicted by
degree distribution alone. This pattern occurs in our 
other network measurements as well. 

The Inet model (Fig.~\ref{fig:degree}(c)) fails to capture the
higher degree (implying additional complexity) in the
periphery. Because the BA model has a minimal degree of three, it is
difficult to compare to the observed networks. However, the decrease of the
$k(\bar{d})$-curves at the largest $\bar{d}$-peak is not conspicuous
in the BA model curves.  Thus, there is no clear core-periphery
dichotomy in the BA model. This too is not surprising, because the BA
model was designed to produce ``scale-free'' networks in the sense of
fractals (if one zooms in on any part of system, it looks similar to the whole).

\begin{figure}
  \resizebox*{0.9 \linewidth}{!}{\includegraphics{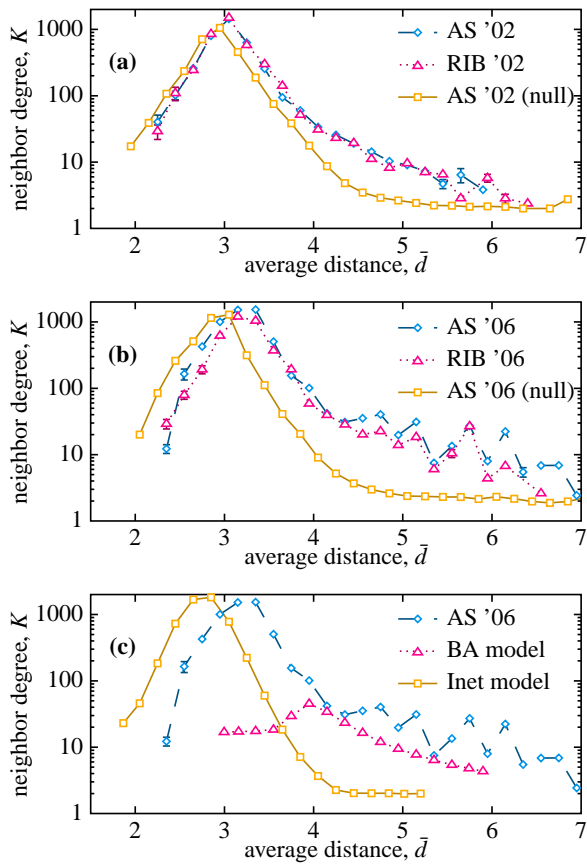}}
  \caption{ Neighbor degree $K$ as a function of the average
    distance $\bar{d}$. The panels and symbols represent the
    same data sets as in Fig.~\ref{fig:density}.}
  \label{fig:nbdeg}
\end{figure}

\subsection{Neighbor degree}

Degree is a property of individual vertices, with no information about how they are interconnected. In this sense degree is a measure of local
network structure. A common way to broaden the perspective to understand
the network's non-local organization~\cite{caida:corr} is to measure the
correlations of degrees between neighbors in the network. There are
three common approaches. The first, known as \textit{assortative mixing
coefficient}~\cite{mejn:rev}, measures the Pearson correlation
coefficient for each edge. This provides one number for the entire
network and is thus appropriate for comparisons between networks. The
second approach makes a density plot that displays the fraction
of edges with degree $(k_1,k_2)$. This kind of two-dimensional plot is
called a \textit{correlation
profile}~\cite{maslov:inet,three:mah}. Correlation profiles provide
more detailed information than the assortative mixing coefficient, but they
are less concise and more sensitive to noisy data. The third approach measures
average neighbor degree
\begin{equation}
  K(i) = \frac{1}{k(i)}\sum_{j\in\Gamma_i} k(j)~,
\end{equation}
(where $\Gamma_i$ is the neighborhood of $i$) as a function of degree
$k(i)$~\cite{pas:inet}. All approaches
must be compared to null models because skewed degree
distributions are known to induce
anti-correlations~\cite{maslov:inet}. The third approach produces a one-dimensional plot
and thus forms a middle ground between the assortative mixing coefficient
and the correlation profile. It is also a method that can
be adapted to our radial-plot framework---by plotting $K$
against $\bar{d}$ we can monitor the correlation between centrality and
neighbor degree. For the AS-level Internet it has been observed
that the $K(k)$-curves decay~\cite{pas:inet}. In other words, high-degree
vertices are, on average, connected to vertices of low degree and vice
versa. Then, since degree decreases with $\bar{d}$,
one would then expect $K$ to be an increasing function
of $\bar{d}$.

As seen in Fig.~\ref{fig:nbdeg}, vertices at intermediate distances
have neighbors of highest degree. The peak in $K(\bar{d})$ coincides
with the largest peak in the histograms found in
Fig.~\ref{fig:density}, and the change of slope in
Fig.~\ref{fig:degree}.  This suggests that the periphery is composed
of two levels: the intermediate majority which is primarily connected
to the core, and the extreme periphery that is connected to other
periphery vertices.

It is also apparent in Fig.~\ref{fig:nbdeg}(a) and (b) that the
null-model qualitatively has the same shape as the observed network; but,
just as for $k$; $K$ are larger in the observed networks than the
null-model.  Also, the Inet model underestimates the average neighbor
degree in the periphery. Finally, the BA model exhibits less
correlation between $K$ and $\bar{d}$.

\begin{figure}
  \resizebox*{0.9 \linewidth}{!}{\includegraphics{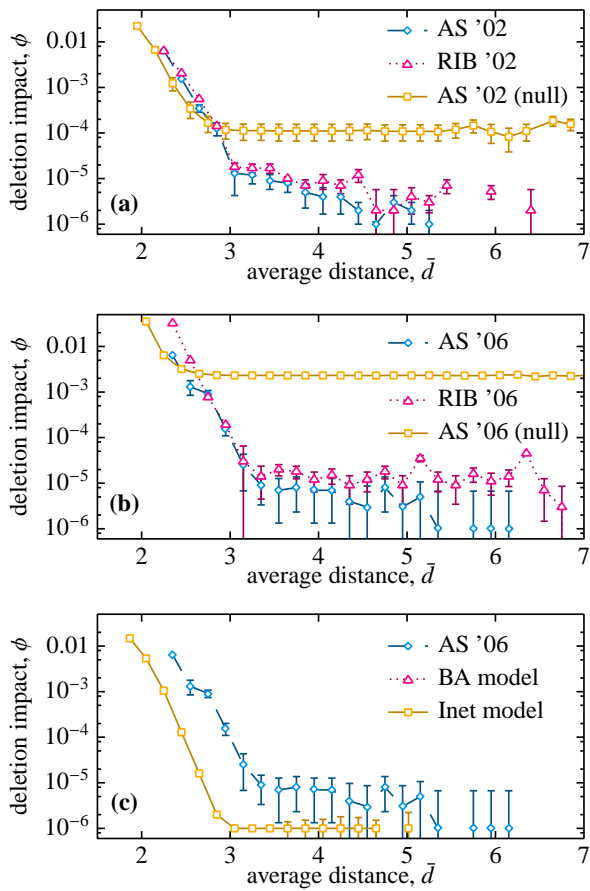}}
  \caption{ Deletion impact $\phi$ as a function of the average
    distance $\bar{d}$. The panels and symbols represent the
    same data sets as in Fig.~\ref{fig:density}. }
  \label{fig:delete}
\end{figure}

\subsection{Deletion impact}
\label{sec:delete}
If a vertex is not actively routing packets due to fault or attack,
other vertices might be affected. We are interested in knowing how susceptible
a given network structure is to random node failures. Assuming that
the network is connected, let $S_i$ be the number of vertices in the
largest connected subgraph after the deletion of $i$.  We define the
\textit{deletion impact} as
\begin{equation} \label{eq:del}
  \phi(i) = \frac{N-1-S_i}{N-2}.
\end{equation}
This measure can take values in the interval $[0,1]$. A value of $0$
means that the entire network, except $i$, is still connected after
the deletion. A value of $1$ means that all of the network's edges
were attached to $i$ and that all of the vertices are isolated after
the deletion.


Fig.~\ref{fig:delete} plots deletion impact as a function of the
average distance for the same data sets as the previous figures. All
curves are roughly decreasing. This means that the
network is more sensitive to the deletion of central, than peripheral,
vertices. This observation is anticipated from earlier studies
showing that the Internet is vulnerable to targeted attacks at the
vertices of highest degree~\cite{alb:attack}  but robust to random
failures. This is because the majority of vertices have low
$\phi$-values.  However, the deletion impact measure can detect more
subtle effects in the periphery.

The first peak in the $\bar{d}$-distribution is, as mentioned above,
around $\bar{d}=3$. At this distance $\phi$ has decreased a thousand
times from the core where $\phi\sim 10^{-2}$. In this quantity we see
a substantial difference from the null-model; the peripheral vertices
of the inferred networks have significantly lower deletion impact than
the peripheral vertices of the null-model networks. This, we believe, is another effect of the high degree of peripheral vertices. The
fact that the periphery is relatively highly connected suggests that
there are alternate routes that could be used if a regular path is
obstructed by a vertex failure. In the case
of the Inet model, which has very few vertices of high $\bar{d}$, the
peripheral $\phi$ values are quite low because the periphery is well
connected to the core. As expected, $\phi=0$ for all vertices in the
BA model since all vertices have degree of at least three. The BA model thus produces
network that are more robust to vertex deletion than the observed networks are.

\begin{figure}
  \resizebox*{0.9 \linewidth}{!}{\includegraphics{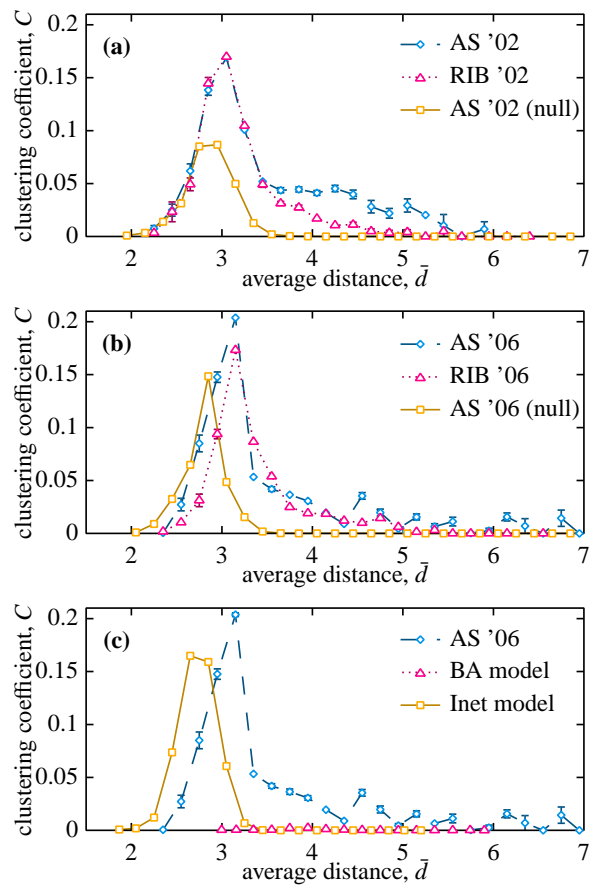}}
  \caption{ Clustering coefficient $C$ as a function of the average
    distance $\bar{d}$. The panels and symbols represent the
    same data sets as shown in Fig.~\ref{fig:density}.}
  \label{fig:clust}
\end{figure}

\subsection{Clustering coefficient}

The \textit{clustering
  coefficient} $C(i)$~\cite{wattsstrogatz} is another frequently studied
network property:
\begin{equation}
  C(i) = M(\Gamma_i)\Big/\dbinom{k(i)}{2}
\end{equation}
$M(X)$ denotes the number of edges in a subgraph $X$. The
clustering coefficient measures how interconnected the neighborhood of
a vertex is. One interpretation is that $C(i)$ is the number of
connected neighbor pairs rescaled by the theoretical maximum.  $C(i)$ can
also be seen as the fraction of triangles that $i$ is a member of, normalized
to the interval $[0,1]$.

In Fig.~\ref{fig:clust} we display the clustering
coefficient as a function of the average distance. The curves for the
observed graph, null-model, and Inet model networks show a peak around the
same point as the peak in the $\bar{d}$-distribution. However, the
null-models do not exhibit as high a degree of clustering in the
periphery as the inferred networks. In other words, there are more triangles in
the periphery than can be expected from only the network's degree
distribution. In fact, for 100 null-model networks based on the AS
'06 network, no triangles existed for $\bar{d}>3.8$ with any vertex
having $\bar{d}>3.8$. This should be compared with 1124 triangles for
the AS '06 network itself (there are even 83 triangles where all
vertices have $\bar{d}>3.8$). This further suggests that
the periphery of the observed AS graphs is complex. As
triangles represent redundancy (the three vertices will still be
connected if any one of the edges are cut) this could help to explain
the increased robustness to deletion seen in Section~\ref{sec:delete}. As seen in
Fig.~\ref{fig:clust}(b), neither the Inet, nor the BA model predict a
significant number of peripheral triangles. The low deletion impact
values for peripheral vertices in these models may be
attributed to the presence of longer cycles.

\begin{figure}
  \resizebox*{0.9 \linewidth}{!}{\includegraphics{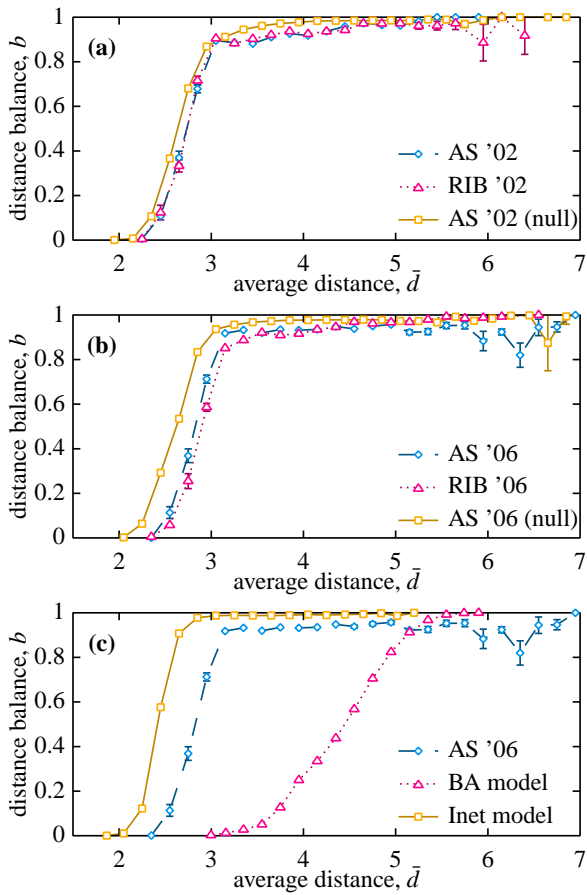}}
  \caption{ Distance balance $b$ as a function of the average
    distance $\bar{d}$. The panels and symbols represent the
    same data sets as shown in Fig.~\ref{fig:density}.}
  \label{fig:balance}
\end{figure}

\subsection{Distance balance}

In the context of scientific collaboration networks it has been
shown~\cite{mejn:scicolpre2} that the number of shortest paths
leaving a vertex via a specific neighbor is skew distributed. In
other words, most of the shortest paths from a vertex $i$ to the rest
of the network traverse a single neighbor of $i$. To rephrase this in
terms of the average distance, central
vertices are likely to have few neighbors with smaller
$\bar{d}$ values. This leads us to another view of centrality. Let the
\textit{distance balance} of $b(i)$ be the fraction of $i$-neighbors $j$ with  $\bar{d}(j)<\bar{d}(i)$. Clearly one can expect this to be an
increasing function of $\bar{d}$, but is it a linear increase?

In Fig.~\ref{fig:balance} we plot the distance balance as a function
of $\bar{d}$. As expected, all of the curves generally increase but
not linearly. Almost all the increase from 0 to 1 takes
place around the highest peak in Fig.~\ref{fig:density}, which gives
another characterization of the core and periphery: in the core, the
typical vertex has relatively few neighbors of higher centrality than
itself (and vice versa in the periphery). The $b(i)$ values in the
peripheral region of all curves approach values close to $1$. In
Fig.~\ref{fig:balance}(b) the curves of the observed data are somewhat
lower. This supports the previous observation that---as seen
previously in quantities such as degree, neighbor degree, and the
clustering coefficient---the periphery is structurally less different
from the core than what can be expected from random networks
constrained to the degree sequence of the observed networks. As seen
in Fig.~\ref{fig:density}(c), the Inet model behaves like the
null-model---the same observation holds for the average neighbor
degree (Fig.~\ref{fig:nbdeg}) and clustering coefficient
(Fig.~\ref{fig:clust}).  Unlike the Inet model, the BA model's curve
increases more smoothly which suggests (in accordance with what has
been observed above) a less pronounced core-periphery structure than
the observed networks.

\section{Summary and conclusions}

This paper investigated how vertex-specific network
measures of the AS level Internet vary with the average distance from
a vertex to the other vertices of the graph. This projection of
vertices to the space of average distances gives a picture of how the network structure changes from the most central to the most peripheral vertices. Using the
distance separation measure we find that there is a well-defined
core-periphery dichotomy in the inferred networks. To some extent
this can be explained as an effect of the set of degrees of the
network---we notice that the average degree as a function of the
average distance has the same qualitative form for the observed
networks as our null-model networks. However, the
periphery is more complex than what is predicted by
degree alone. This is manifested in higher average degree, higher
average neighbor degree, lower deletion impact, higher clustering
coefficient, and lower distance balance than the observed
networks. To summarize, the AS graph has a more clear
split into a core and a periphery than can be anticipated by its
degree distribution and simple models of scale-free networks. At the
same time, the split is less dramatic and more nuanced than expected from a strict hierarchy. The additional network structure in the periphery may have consequences for spread of attacks and methods to defend against attack.
Further, the two topology generators (Inet and
BA model) that we tested could be extended to
model the periphery more accurately.

We used two kinds of observed AS data---easily accessible router RIBs
and more complete data sets where edges missing from the RIBs are
added. The effect of the missing edges is clearly visible: the
peripheries of the RIB-networks (with missing edges) have lower
average degree, lower number of triangles, and other traits. On the
other hand, the missing links do not change the network structure
qualitatively. Our conclusions would be unchanged if we used only the
RIB data.

Future modeling and measuring research needs to be undertaken to
elucidate the detailed structure of the core and periphery of the AS
graph. Furthermore, the structures should be related to the strategies
of AS management~\cite{daub:as,peer:chang,inet}.

\acknowledgements{
  PH acknowledges financial support from the Wenner-Gren
  foundations. The authors acknowledge the support of the
  National Science Foundation (grants CCR--0331580 and CCR--0311686),
  and the Santa Fe Institute.
}

\end{document}